\documentclass[reprint, superscriptaddress, secnumarabic, amssymb, nobibnotes, aps, prl]{revtex4-1}

\usepackage{amsmath}
\usepackage{graphicx}
\usepackage{epstopdf}
\usepackage[]{natbib}
\usepackage{siunitx}

\begin{document}

\title{\textrm{Time-reversal symmetry breaking in noncentrosymmetric superconductor Re$_{6}$Hf}: further evidence for unconventional behaviour in the $\alpha$-Mn family of materials}
\author{D. Singh}
\affiliation{Department of Physics, Indian Institute of Science Education and Research Bhopal, Bhopal, 462066, India}
\author{J. A. T. Barker}
\affiliation{Laboratory for Muon Spin Spectroscopy, PSI, CH-5232 Villigen PSI, Switzerland}
\author{A. Thamizhavel}
\affiliation{Department of Condensed Matter Physics and Materials Science, Tata Institute of Fundamental Research, Mumbai 400005, India}
\author{D. McK. Paul}
\affiliation{Physics Department, University of Warwick, Coventry CV4 7AL, United Kingdom}
\author{A. D. Hillier}
\affiliation{ISIS Facility, STFC Rutherford Appleton Laboratory, Harwell Science and Innovation Campus, Oxfordshire, OX11 0QX, UK}
\author{R. P. Singh}
\email[]{rpsingh@iiserb.ac.in}
\affiliation{Department of Physics, Indian Institute of Science Education and Research Bhopal, Bhopal, 462066, India}

\date{\today}
\begin{abstract}
\begin{flushleft}
\end{flushleft}
The discovery of new families of unconventional superconductors is important both experimentally and theoretically, especially if it challenges current models and thinking. By using muon spin relaxation in zero-field, time-reversal symmetry breaking has been observed in Re$_{6}$Hf. Moreover, the temperature dependence of the superfluid density exhibits $s$-wave superconductivity with an enhanced electron-phonon coupling. This, coupled with the results from isostructural Re$_{6}$Zr, shows that the Re$_{6}$$X$ family are indeed a new and important group of unconventional superconductors.
\end{abstract}
\maketitle

Superconductivity is a complex phenomenon, and provides a rich experimental and theoretical playground in which to investigate physics. To this date, our microscopic understanding is primarily built upon the theory of Bardeen, Cooper, and Schrieffer (BCS), which describes how conventional superconductivity arises due to the coupling of electrons as spin-singlet Cooper pairs, mediated by the electron-phonon interaction \cite{BCS}. However, there exists whole classes of superconducting materials in which the superconductivity cannot be described by this conventional theory.  It is in these systems that the most exciting, rich, and varied physics can occur, and which offer the most interesting potential applications for the future.

The central principle of BCS theory is the formation of Cooper pairs at the superconducting transition temperature, $T_{c}$. The superconducting ground state is described by a macroscopic wave function representing these Cooper pairs, which behaves like a thermodynamic order parameter.  In order to satisfy the Pauli Exclusion Principle, this Cooper pair wavefunction should be antisymmetric under particle exchange. Thus, pairing states form in either spin-singlet configurations with even parity, or as spin-triplet pairs with odd parity \cite{Bauer2004}. Conventionally, such pairing is only possible in materials that have inversion symmetry, i.e. centrosymmetric structures. Noncentrosymmetric materials lacking inversion symmetry exhibit a non-uniform lattice potential, which leads to an antisymmetric spin-orbit coupling (ASOC) \cite{EI,rashba}. This breaks the degeneracy of the conduction band electrons and may cause the superconducting pair wave function to form in a mixed-parity superconducting state \cite{TP}. The mixed-pairing state can lead noncentrosymmetric superconductors (NCS) to exhibit significantly different properties from conventional superconducting systems, e.g. nodes in the superconducting gap [2], upper critical fields exceeding the Pauli limiting field \cite{PA}, time-reversal symmetry breaking (TRSB), and very recently the signature of topologically protected zero-energy surface- or edge states \cite{GB}.

TRSB remains an exceptionally rare phenomenon in this class of materials. It was concluded that only certain irreducible representation of the crystal point group will permit the pairing states with TRSB, whereas pairing states constructed from other irreducible representations, also with mixed parity, would not necessarily be expected to exhibit TRSB. This makes the occurrence of TRSB in NCS much more interesting. To date, the only NCS that have yet been reported to show TRSB are LaNiC$_{2}$ \cite{ADJ}, Re$_{6}$Zr \cite{RPS}, locally noncentrosymmetric SrPtAs \cite{PKB} and La$_{7}$Ir$_{3}$ \cite{JAT}. Meanwhile, TRS is found to be preserved in several other NCS including Ca(Ir,Pt)Si$_{3}$ \cite{RAD}, La(Rh,Pt,Pd,Ir)Si$_{3}$ \cite{VK,MS,VKD}, Mg$_{10}$Ir$_{19}$B$_{16}$ \cite{TKF}, Re$_{3}$W \cite{PAM} and Mo$_{3}$Al$_{2}$C \cite{EC}. The results obtained from transverse field muon measurements seem to imply a dominant $s$-wave component in the superconductivity of these systems, however the full story requires a good understanding of the point group symmetry of the materials in question. It is therefore of great importance to study new noncentrosymmetric structures, to understand the presence and absence of TRSB in NCS.
 
Muon spin rotation and relaxation ($\mu$SR) is one of the most direct ways of confirming the presence of an unconventional superconducting state \cite{AS,SLL,AY}. The magnetic penetration depth can be accurately determined using $\mu$SR, and measuring the temperature dependence of this quantity yields information about the symmetry of the superconducting gap. This technique is also very sensitive to the tiny magnetic signal that may be associated with the formation of spin-triplet Cooper pairs. This feature can also be used to unambiguously establish broken time-reversal symmetry in the superconducting state \cite{MSK}.

In this paper, we report evidence for broken time reversal symmetry (TRS) in the superconducting state of the binary transition metal compound Re$_{6}$Hf, which is isostructural to Re$_6$Zr. Our results show that the TRSB signal in Re$_6$Hf is very similar to the signal observed in Re$_6$Zr, and shows that Re$_6X$ is an important family of unconventional superconducting materials, in which the effect of spin-orbit coupling on the superconducting ground state can be investigated.

The preparation of the polycrystalline Re$_{6}$Hf sample used in this work is described in Ref.~\cite{DSA}. Powder X-ray diffraction (XRD) data confirms that the sample has the $\alpha$-$\textit{Mn}$ crystal structure (space group $I \bar{4}3m$ No. 217), with no impurity phases detected to within the sensitivity of the XRD technique. Magnetization, heat capacity, and $\mu$SR measurements indicate that Re$_{6}$Hf is a bulk superconductor with $T_{c}=5.98 \pm 0.02$ K and that the superconducting ground state appears to be predominantly s-wave with enhanced electron-phonon coupling \cite{DSA}. The MuSR instrument at the ISIS pulsed muon and neutron spallation source was used to carry out the $\mu$SR measurements. A detailed account of the $\mu$SR technique may be found in Ref.~\cite{SLL}.  Stray fields at the sample position due to neighboring instruments and the Earth's magnetic field are canceled to within $\sim$1.0 $\mu$T using three sets of orthogonal coils and an active compensation system.  The powdered Re$_{6}$Hf sample was mounted on a silver holder and placed in a sorption cryostat, which operated in the temperature range 0.3 K - 10 K.

\begin{figure}
\includegraphics[width=1.0\columnwidth]{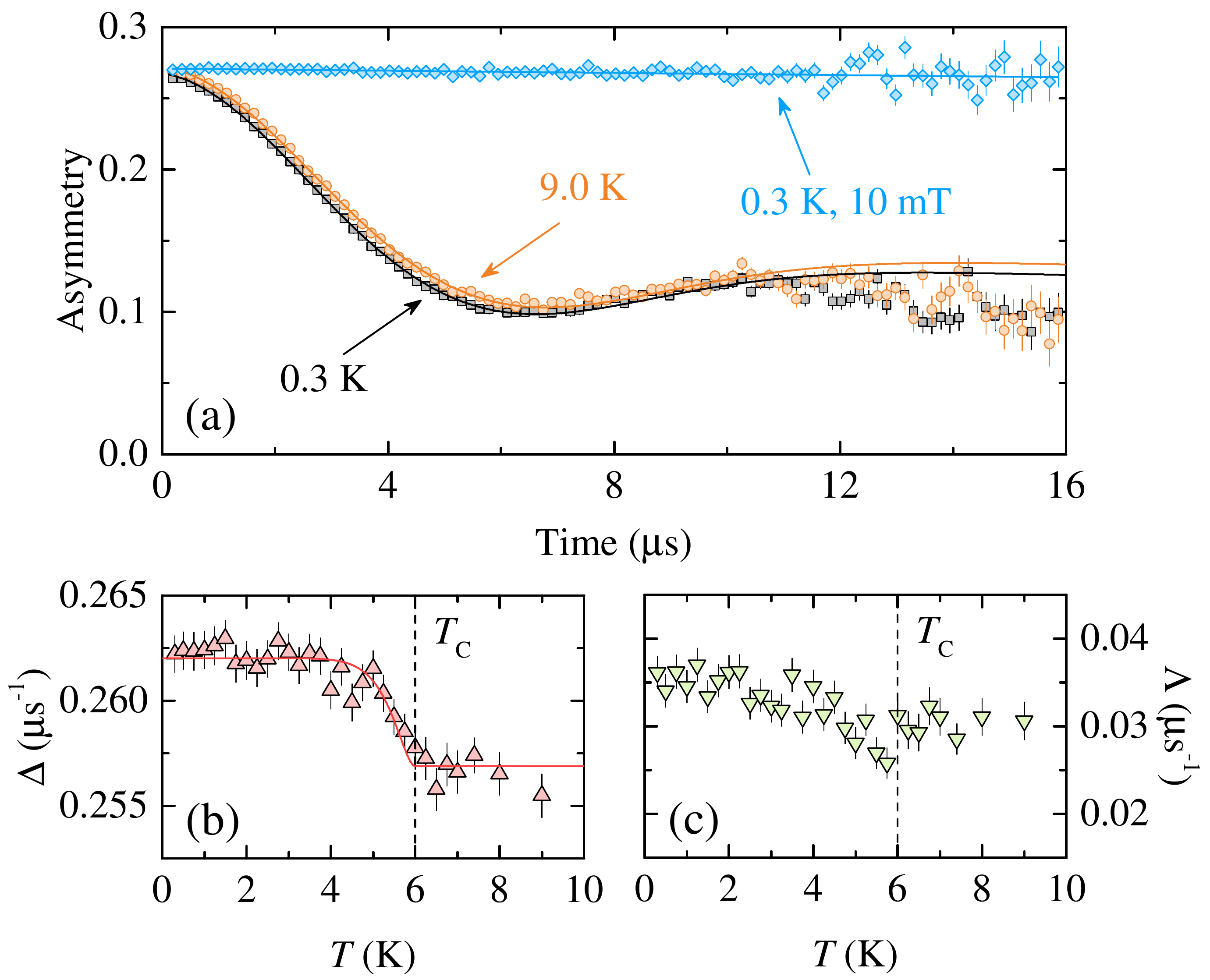}
\caption{\label{Fig2:ZFM} (a) ZF- and LF-$\mu$SR spectra collected above (circles) and below (squares) $T_{c}$ with least-squares fits using the model of Eq.2 (solid lines).  In ZF there is a clear difference between the spectra, indicating the presence of spontaneous fields in the superconducting state.  The effect of applying a small LF field of 10 mT is also shown (diamonds). (b) Temperature dependence of the Gaussian relaxation parameter, $\Delta$, which clearly shows spontaneous fields appearing below $T_{c}$ $\sim$ 6.0 K. (c) The electronic relaxation rate, $\Lambda$, exhibits a weak temperature dependence, with no appreciable change at $T_c$.}
\end{figure}

Zero-field (ZF) muon-spin relaxation data were collected at several temperatures above and below $T_c$. Figure 1(a) shows typical asymmetry spectra collected at 9.0 K and 0.3 K. There is a subtle change in the relaxation behavior upon lowering the temperature below $T_c$. No oscillatory components are present in the spectra, and no loss of asymmetry is observed, which rules out the presence of spontaneous coherent magnetic fields associated with magnetic order. In the absence of atomic moments, and at low temperatures where muon diffusion is not appreciable, the depolarization of the muon is due to the presence of static, randomly oriented nuclear moments. This depolarization can be modeled by the Kubo-Toyabe equation \cite{RSH}
\begin{equation}
G_{\mathrm{z}}(t) = \frac{1}{3}+\frac{2}{3}(1-\Delta^{2}t^{2})\mathrm{exp}\left(-\frac{1}{2}\Delta^{2}t^{2}\right) ,
\label{eqn41:zf}
\end{equation}
where $\Delta$ measures the width of the nuclear dipolar field experienced by the muons. The ZF asymmetry spectra are well described by the function
\begin{equation}
A_\mathrm{ZF}(t) = A_{0}G_{\mathrm{z}}(t)\mathrm{exp}(-\Lambda t)+A_{\mathrm{BG}} ,
\label{eqn2:tay}
\end{equation}
where $A_{0}$ and $A_{\mathrm{BG}}$ are the sample and background asymmetries, respectively.  The exponential relaxation rate, $\Lambda$, is typically associated with dynamic electronic moments, which fluctuate on a timescale that is much faster than the muon lifetime. Above model fits poorly above $10\mu\mathrm{s}$. It might be due to an additional contribution, or the exact model of the Kubo-Toyabe is not correct in the present case. 

The temperature dependence of the fit parameters $\Delta$ and $\Lambda$ are presented in Fig.~1(b) and Fig.~1(c), respectively.  The sample and background asymmetries have the approximately temperature independent values $A_0 = 0.1748(8)$ and $A_{bg} = 0.0963(3)$. The Gaussian relaxation rate, $\Delta$, shows a systematic increase below the superconducting transition temperature, whereas $\Lambda$ exhibits only a very weak temperature dependence, with no discernible change in behaviour at $T_c$.

The shape of $\Delta(T)$ has been empirically modelled by a generic $N$-fluid equation of the form $\Delta(T)/\Delta(0)=1-(T/T_c)^n$ - shown as the solid line in Fig. 1(b).  $\Delta$ increases from a baseline level of $0.2570(3)~\mu\mathrm{s}^{-1}$ to a low temperature value of $0.2621(3)~\mu\mathrm{s}^{-1}$. $T_c$ was kept fixed to 6 K for this procedure, with the power $n=11(2)$.  This large value for $n$ highlights the rapidity with which the signal develops at $T_c$.  We note that the high temperature value of $\Delta$ is very close to that measured in Re$_6$Zr ($0.256~\mu\mathrm{s}^{-1}$), which is a good indication that Re nuclear moment is the primary contributor to the nuclear dipolar field.

The nature of the relaxation signal can be further probed by the application of a longitudinal field (LF), as this allows the magnitude of the internal field to be estimated.  When $\mu_0H_\mathrm{app}\approx 10B_\mathrm{int}$, the muon spins become fully decoupled from a static internal field - experimentally we observe this as a flat asymmetry spectra.  This is exactly what we see upon application of a small LF of 10 mT, as depicted by the blue diamonds in Fig. 1(a).  This shows that the magnitude of the magnetic signal is $\le 1$ mT. In fact, the mode (or peak) of the internal field distribution, $|B_\mathrm{int}|$ is related to $\Delta$ by the equation
\begin{equation}
\label{eq:bint}
	|B_\mathrm{int}|=\sqrt{2}\frac{\Delta}{\gamma_\mu},
\end{equation}
where $\gamma_\mu/2\pi=135.5~$MHz/T is the muon gyromagnetic ratio.  Subtracting the high temperature value of $\Delta$ from the low temperature value yields the change $\delta\Delta=0.0052(1)~\mu\mathrm{s}^{-1}$, which together with Eq.~\eqref{eq:bint} implies that the peak value of the local field distribution below $T_c$ increases by $\delta B_\mathrm{int}=0.086~$G.  

The structure of the superconducting gap in Re$_6$Hf was determined by measuring the superfluid density using transverse field $\mu$SR (TF-$\mu$SR).  A magnetic field of 25 mT was applied above $T_{c}$, before cooling through the superconducting transition to a temperature of 300 mK, in order to stabilize the flux line lattice in the mixed state of the superconductor. Typical muon asymmetry signals are presented for temperatures above and below $T_{c}$ in Fig.~2(a).  Above $T_c$, the spectra oscillate with a frequency that corresponds to Larmor precession of the muon spin in the applied field, damped with a weak Gaussian relaxation that is due to the nuclear dipolar field.  Below $T_c$, this Gaussian damping becomes more pronounced, due to the development of the field distribution associated with the vortex lattice.  The time evolution of the asymmetry is described well by a sum of cosines, each damped with a Gaussian relaxation term:
\begin{equation}
A_\mathrm{TF}(t) = \sum_i^n A_{i}\exp\left(-\frac{1}{2}\sigma_i^2t^2\right)\cos(\gamma_\mu B_it+\phi).
\label{eqn3:Tranf}
\end{equation}
In this equation, $A_i$, $\sigma_i$, and $B_i$ correspond to the asymmetry, depolarization rate, and field of the $i$'th oscillating component, $\phi$ is a shared phase offset, and $\gamma_{\mu}/2\pi$ = 135.5 MHz/T is the muon gyromagnetic ratio.  To fit the low temperature data adequately, $n=3$ oscillating components were required. Above $T=4$ K, the signal from one of the components had become negligibly small, and the number of oscillations was reduced to $n=2$.  The depolarization rate of the $i=1$ oscillating component was fixed to $\sigma_1=0$ throughout the fitting procedure, which accounts for muons stopping in the silver sample holder as they do not appreciably depolarize over the time-scale of the experiment. Furthermore, the value of $A_1=0.0871$ determined from the lowest temperature measurement was kept fixed.

\begin{figure}
\includegraphics[width=1.0\columnwidth]{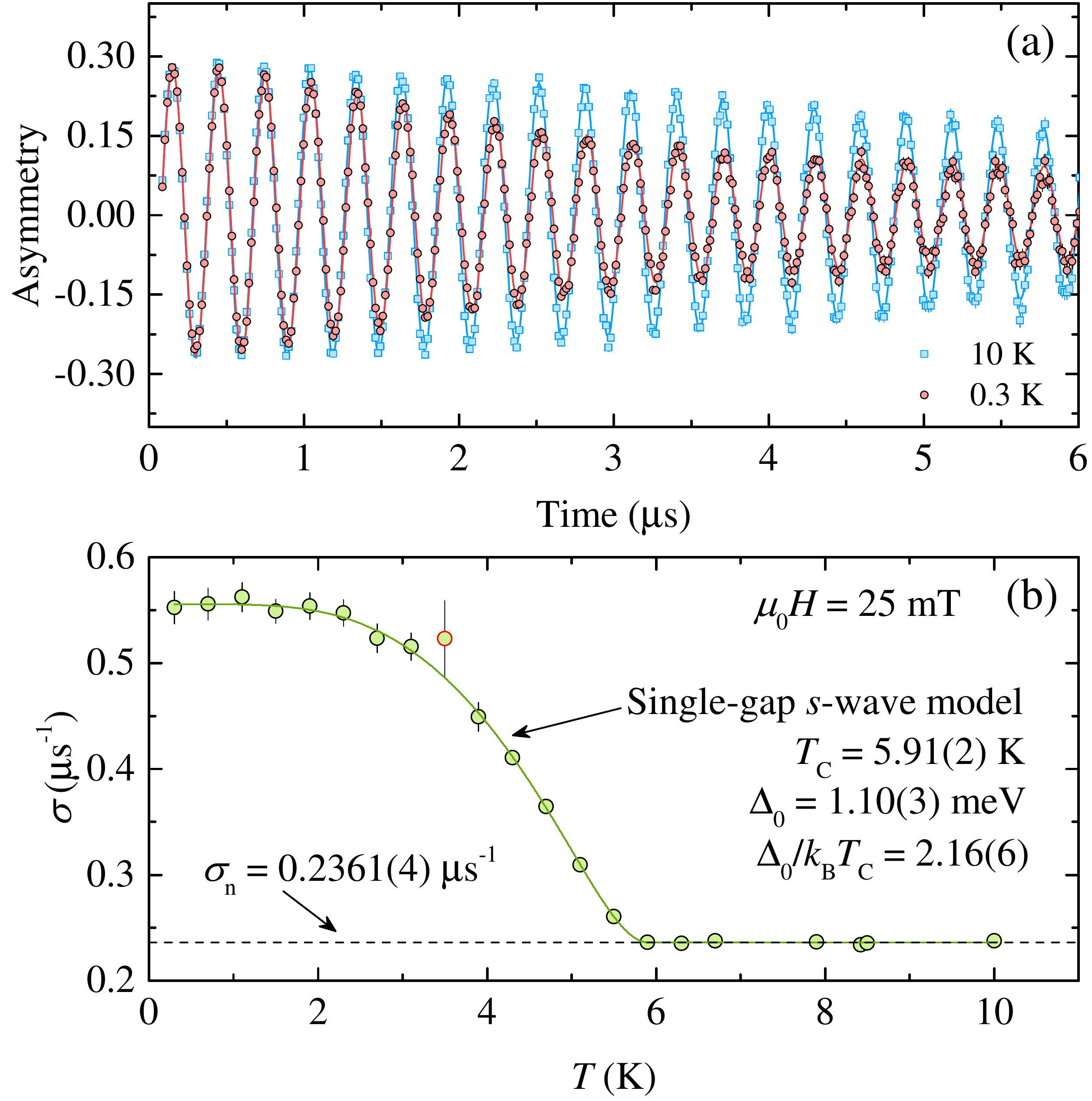}
\caption{\label{Fig3:TF} (a) Representative TF - $\mu$SR signals collected at (a) 300 mK and (b) 10 K in an applied magnetic field of 25 mT.  The solid lines are fits using Eq. 3. The effect of the flux line lattice can be seen in the top panel as the strong Gaussian decay envelope of the oscillatory function.  Above $T_{c}$, the depolarization is reduced and is due to the randomly oriented array of nuclear magnetic moments. (b) Temperature dependence of the TF- $\mu$SR depolarization rate collected in a field of 40 mT.  The data point with the red outline has been masked from the fitting as an outlier.}
\end{figure}

The second-moment method, described in Ref.~\cite{MKSK}, was used to calculate the total depolarization rate, $\sigma$. The temperature dependence of $\sigma$ is displayed in Fig. 2(b).  One anomalous data point has been marked, and excluded from further analysis. Above $T_c$, $\sigma$ is temperature independent, and has the average value $\sigma_\mathrm{n}=0.2361(4)~\mu\mathrm{s}^{-1}$.  We associate this with relaxation due to the nuclear dipolar field.  As the temperature is lowered through $T_c$, $\sigma$ begins to increase, which corresponds to the development of the field distribution associated with the vortex lattice.  This field distribution has associated with it a Gaussian damping rate, $\sigma_\mathrm{vl}$, and this adds in quadrature with $\sigma_\mathrm{n}$ to produce the total depolarization rate:
\begin{equation}
\sigma^{2} = \sigma_{\mathrm{n}}^{2}+\sigma_{\mathrm{vl}}^{2}.
\label{eqn4:sigma}
\end{equation}
Below about 1.5 K, $\sigma$ is once again temperature independent, which suggests that the superconductivity should be well described by a fully-gapped, $s$-wave model.

For a single-gap $s$-wave superconductor in the dirty limit, the temperature dependence of the superfluid density, $\rho_\mathrm{s}$, is given by
\begin{equation}
\rho_\mathrm{s} = \frac{\lambda^{-2}(T)}{\lambda^{-2}(0)} = \frac{\Delta(T)}{\Delta_0}\mathrm{tanh}\left[\frac{\Delta(T)}{2k_{B}T}\right],
\label{eqn5:lpd}
\end{equation}
where $\Delta(T)$ is the BCS approximation for the temperature dependence of the superconducting gap energy, and $\Delta_0$ is the low temperature value of the gap energy.  The dirty limit is used when the BCS coherence length, $\xi_{0}$, is of a similar order of magnitude to the electronic mean free path, $\textit{l}$.  The justification for using the dirty limit arises from a calculation of $\xi_{0}$/$\textit{l}$ = 1.08, based on the experimental data in Ref.~\cite{DSA}. 
For high-$\kappa$ superconductors in the low field limit, a simple numerical prefactor relates $\sigma_{\mathrm{vl}}$ and the inverse squared magnetic penetration depth, $\lambda^{-2}$, i.e. $\sigma_{\mathrm{vl}}$ $\propto$ $\lambda^{-2}$. Therefore, $\sigma_{\mathrm{vl}}$ is directly related to the superfluid density by the equation
\begin{equation}
\frac{\sigma_{\mathrm{vl}}(T)}{\sigma_{\mathrm{vl}}(0)} = \frac{\lambda^{-2}(T)}{\lambda^{-2}(0)}
\label{eqn6:sfd}
\end{equation}

Combining eqs. (5)-(7) produces a model for $\sigma$, with the fit parameters $\Delta_0$, $\sigma_\mathrm{vl}(0)$, $\sigma_\mathrm{n}$, and $T_c$.  A fit to this model is shown by the solid line in Fig. 2(b).  The fitted value for the transition temperature, $T_c=5.91(2)$ K, is in good agreement with the bulk measurements of $T_c=5.98$ K.  The energy gap has a maximum magnitude of $\Delta_0=1.10(2)$ meV, which expressed as a ratio of $T_c$ and the Boltzmann constant has the value $\Delta_0/k_\mathrm{B}T_c=2.16(4)$.  This is much larger than the BCS expectation (1.764), and is an indication that the coupling strength is stronger than the conventional mechanism in this material. This result is paralleled in Re$_6$Zr, in which a similar measurement using muons determined that $\Delta_0/k_\mathrm{B}T_c=2.1(2)$.  This suggests that the pairing mechanism in this family of materials is similar.  Finally, the low temperature value of the depolarization rate is $\sigma_\mathrm{vl}(0)=0.503(3)~\mu\mathrm{s}^{-1}$.  The zero-temperature value of $\lambda$ can be estimated from this value of $\sigma_{\mathrm{vl}}$(0). As mentioned previously, a simple numerical prefactor relates $\lambda$ to $\sigma_{\mathrm{vl}}$ via the equation
\begin{equation}
\frac{\sigma_{\mathrm{vl}}^2(0)}{\gamma_{\mu}^2} = 0.00371 \frac{\Phi_{0}^{2}}{\lambda^{4}(0)} ,
\label{eqn7:lam}
\end{equation}
where $\Phi_{0}$ is the magnetic flux quantum. This gives $\lambda$(0) = 462(1) nm, which is slightly larger than the value found in Ref.~\cite{DSA}.
	
	It should be noted that in the case of dirty limit superconductors there is a possibility that the actual temperature dependence of the superfluid density might be smeared out due to scattering from defects or impurities. This particular remark was also made in CaIrSi$_{3}$ \cite{BAFS}, where they have comprehensively discussed the implications of polycrystalline sample on data interpretation of TF-$\mu$SR results. It was shown that majority of the NCSs with unconventional pairing symmetry were in the clean limit and the dirty limit superconductivity may result in the loss of information about the true pairing symmetry. Therefore, it is highly desirable to do the measurements on a high quality single crystals of Re$_{6}$Hf in order to know the true behavior of superfluid density and inspect for the possibility of non-s wave gap symmetries.\\  
The results of the ZF-$\mu$SR study have shown that the magnetic signal below $T_c$ is weak, quasi-static with respect to the muon lifetime, and switches on rapidly at the superconducting transition.  We therefore conclude that time-reversal symmetry is broken in the superconducting state in Re$_6$Hf, adding to the other NCS with TRSB: LaNiC$_{2}$, Re$_{6}$Zr, and La$_{7}$Ir$_{3}$ (and the locally non-centrosymmetric SrPtAs). In LaNiC$_2$, all of the allowed states with TRSB correspond to non-unitary triplet pairing, in which the Cooper pairs are spin polarized. However, all of these gap functions exhibit nodes, and all experimental measurements of the superfluid density indicate fuly-gapped superconductivity. Furthermore, introducing a significant spin-orbit coupling means that these states are no longer allowed to form at $T_c$. This seems to suggest that the TRSB in LaNiC$_2$ occurs in spite of the fact that the crystal structure is noncentrosymmetric, and that a significant spin-singlet/triplet admixture ground state is forbidden.  Recent measurements of the gap structure in LaNiC$_2$ have revealed two-gap superconductivity, and it was recently proposed that the superconducting ground state is fully gapped, with non-unitary triplet pairing between electrons that exist on different orbitals \cite{ZFW}.

The point group analysis of Re$_6$Zr \cite{RPS} (which is isostructural to Re$_6$Hf with point group $T_d$) shows that a mixed singlet/triplet state that breaks TRS is allowed, due to the higher crystallographic symmetry in these materials.  $\mu$SR measurements of the superfluid density are well described by a single gap,, consistent with an $s$-wave pairing symmetry. However, recent point-contact Andreev reflectometry measurements have found two superconducting gaps in Re$_6$Zr \cite{PP}. In this situation, if the interaction on the same Fermi surface (FS) pocket is less than the interaction between two different FS pockets, then the interaction will be dominated by the Coulomb repulsion. The superconducting instability will correspond to a 2D representation of the cubic point group where the time-reversal symmetry will be broken \cite{ZFC}. In this case, the superconductivity is conventional, yet exhibits TRSB. A second explanation could be that of the mechanism governing superconductivity in LaNiC$_2$ and LaNiGa$_2$ \cite{ZFW}.

Another thing to consider is the comparative effect of spin-orbit coupling between Re$_6$Zr and Re$_6$Hf.  Since Hf is more massive than Zr, the spin-orbit coupling is expected to be stronger, as it increases with $Z^4$.  However, the TRSB signal observed in ZF in Re$_{6}$Zr is remarkably similar to that seen in Re$_{6}$Hf - The magnitude and shape of the increase in signal from the muon results is nearly identical.  This indicates that the mechanism for superconductivity in both these materials is the same, and that the increased spin-orbit coupling does not lead to an increase in the strength of the spin-triplet channel. The conclusion is that, while ASOC is required in order to allow spin-singlet/triplet mixing, it does not affect the underlying electronic characteristics of the spin-triplet channel, which exists regardless of whether a strong ASOC is present.

In conclusion, we have determined that the superconducting ground state in Re$_{6}$Hf breaks time reversal symmetry. However, the TF data suggest that the superconducting order parameter is described well by an isotropic gap with $s$-wave pairing symmetry and enhanced electron-phonon coupling, similar to that of Re$_{6}$Zr.  The current results suggest a complex superconducting ground state of a dominant $s$-wave component with a smaller triplet component. Further experimental work on single crystals, coupled with theoretical work, is required to fully determine the nature of the superconductivity in this important family of materials.

\section{Acknowledgments}

R.~P.~S.\ acknowledges Science and Engineering Research Board, Government of India for the Ramanujan Fellowship through Grant No. SR/S2/RJN-83/2012, YSS/2015/001799 and Newton Bhabha funding. We thank ISIS, STFC, UK for the muon beamtime to conduct the $\mu$SR experiments.

\end{document}